\documentstyle[amstex,aps,epsfig]{revtex}

\newtheorem{theorem}{Theorem}
\newtheorem{acknowledgement}[theorem]{Acknowledgement}

\begin{document}
\title{Spinodal Instabilities and the Dark Energy Problem}
\author{D. Cormier$^{\left( a\right) }$, R. Holman$^{\left( b\right) }$}
\address{$^{\left( a\right) }$ Institute for Physics, University of Dortmund, 
\\ D-44221 Dortmund, Germany\\
$^{\left( b\right) }$ Physics Department, Carnegie Mellon University,\\
Pittsburgh PA\ 15213, U.S.A.}
\date{\today}
\maketitle

\begin{abstract}
The accelerated expansion of the Universe measured by high redshift Type Ia
Supernovae observations is explained using the non-equilibrium dynamics of 
{\em naturally} soft boson fields. Spinodal instabilities inevitably present
in such systems yield a very attractive mechanism for arriving at the
required equation of state at late times, while satisfying all the known
constraints on models of quintessence.
\end{abstract}

\bigskip

One of the most startling developments in observational cosmology is the
mounting evidence for the {\em acceleration} of the expansion rate of the
Universe\cite{SNIa}. Coupled with cluster abundance and CMB observations,
these data can be interpreted as evidence for a cosmological constant $%
\Lambda $ which contributes an amount $\Omega _{\Lambda }\simeq 0.7$ to the
critical energy density while matter contributes $\Omega _{\text{matter}%
}\simeq 0.3$, leading to a flat FRW\ cosmology (see \cite{Bahcalletal} and
references therein).

While introducing a cosmological constant may be a cosmologically sound
explanation of the observations, it is a worrisome thing to do indeed from
the particle physics point of view. It is hard enough to try to explain a
vanishing cosmological constant, given the various contributions from
quantum zero point energies, as well as from the classical theory\cite
{CosmoConstRev}, but at least one could envisage either a symmetry argument
(such as supersymmetry, if it were unbroken) or a dynamical approach (such
as the ill-fated wormhole approach\cite{wormholes}) that could do the job.
It is much more difficult to see how cancellations between all possible
contributions would give rise to a {\em non-zero} remnant of order $10^{-47}$
GeV$^{4}$ which is {\em extremely} small compared to $M_{\text{Planck}}^{4}$
or $M_{\text{SUSY}}^{4}$, the ``natural'' values expected in a theory with
gravity or one with a supersymmetry breaking scale $M_{\text{SUSY}}$. Even
from the cosmological perspective, a cosmological constant begs the question
of why its effects are dominating {\em now}, as opposed to any time prior to
today, especially given its different redshifting properties compared to
matter or radiation energy density.

These fine tuning problems can at least be partially alleviated if instead
of using a ${\em constant}$ energy density to drive the accelerated
expansion, a dynamically varying one were used instead. This is the idea
behind {\it Quintessence} models\cite{Quintessence}. A\ scalar field whose
equation of state violated the strong energy condition (i.e. with $\rho +3P<0
$ ) during its evolution\cite{scalarfieldcosm} would serve just as well as a
cosmological constant in terms of explaining the data, as long as its
equation of state satisfied the various known constraints for such theories 
\cite{constraints}. However, an arbitrary scalar field whose energy density
dominates the expansion rate is not sufficient to get out of all fine tuning
problems; in particular, for a field of sufficiently small mass that it
would only start evolving towards its minimum relatively recently (i.e.
masses of order the inverse Hubble radius), the ratio of matter energy
density $\rho _{m}$ to field energy density $\rho _{\phi }$ would be need to
be incredibly fine tuned at early times so as to have $\rho _{m}/\rho _{\phi
}\sim 1$ today. The quintessence approach uses so-called tracker fields\cite
{Tracker} that have potentials that drive the field to attractor
configurations that have a a fixed value of $\rho _{m}/\rho _{\phi }$. Thus,
for these models, regardless of the initial conditions, the intermediate
time value of $\rho _{m}/\rho _{\phi }$ will always be the same. The only
fine tuning required in these models is the timing of the deviation of $%
\rho_\phi$ from the tracking solutions to allow it to satisfy the condition $%
\rho_{\phi} +3P_{\phi}<0$ today. 

What we propose is a working alternative to the idea of tracker fields
without the usual fine tuning problems. Recent work on the non-equilibrium
dynamics of quantum fields\cite{NEQFT,CormierThesis} has shown that under
certain circumstances the back reaction of quantum fluctuations can have a
great influence on the evolution of the quantum expectation value of a field
\cite{LargeN,Spinodal}, to the extent that using the classical equations of
motion can grossly misrepresent the actual dynamics of the system. What we
will show below is that we can make use of this modified dynamical behavior
to construct models that might allow for a more natural setting for a late
time cosmological constant.

The class of models we consider are those using pseudo-Nambu-Goldstone
bosons (PNGBs) to construct theories with naturally light scalars\cite
{HillRoss}. Such models have been used for late time phase transitions\cite
{LateTimePT}, as well as to give rise to a cosmological ``constant'' that
eventually relaxed to zero\cite{FriemanHillStebWaga}, not unlike what we
want to do here. However, our take on these models will be significantly
different from that of ref.\cite{FriemanHillStebWaga}.

We can write the required energy density as $\rho _{\text{Dark Energy}}\sim
\left( 10^{-3} \text{ eV}\right) ^{4}$, which is suggestive of a light
neutrino mass scale\cite{superK1,superK2}. There is a way to construct
models of scalar fields coupled to neutrinos where the scalar field
potential naturally (in the technical sense of t'Hooft\cite{t'Hooft})
incorporates the small mass scale $m_{\nu }^{4}$.

Consider a Lagrangian containing a Yukawa coupling of the form\cite{HillRoss}%
: 
\begin{equation}
-{\cal L}_{\text{Yuk}}=\sum_{j=0}^{N-1}\left( m_{0}+\varepsilon \exp i\left( 
\frac{\Phi }{f}+\frac{2\pi j}{N}\right) \right) \overline{\nu }_{jL}\nu
_{jR}+\text{h.c.\label{Yukawa}}  \label{YukLag}
\end{equation}
The scale $f$ is the scale at which a the global symmetry that gives rise to
the Nambu-Goldstone mode $\Phi $ is spontaneously broken. The Lagrangian $%
{\cal L}_{\text{Yuk}}$ is to be thought of as part of the low-energy
effective theory of $\Phi $ coupled to neutrinos at energies below $f$.

The term proportional to $\varepsilon $ could be obtained by a coupling to a
Higgs field $\chi $ that acquires an expectation value $\left\langle \chi
\right\rangle =f/\sqrt{2}\exp i\frac{\Phi }{f}$. Note that in the absence of 
$m_{0}$ this Yukawa term possesses a continuous chiral $U\left( 1\right) $
symmetry. The term proportional to $m_{0}$ breaks this symmetry explicitly
to a residual discrete $Z_{N}$ symmetry given by: 
\begin{equation}
\nu _{j}\rightarrow \nu _{j+1},\text{ }\nu _{N-1}\rightarrow \nu _{0},\text{ 
}\Phi \rightarrow \Phi +\frac{2\pi f}{N}.  \label{ZNsym}
\end{equation}
This interaction can generate an effective potential for the Nambu-Goldstone
mode $\Phi $ which must vanish in the limit that $m_{0}\rightarrow 0$ which
is equivalent to the vanishing of the neutrino masses. Since $\Phi $ is an
angular degree of freedom, it should not be a surprise that the effective
potential is periodic and of the form 
\begin{equation}
V\left( \Phi \right) =M^{4}\left( 1+\cos \frac{N\Phi }{f}\right) .
\label{PGNBeffpot}
\end{equation}
Here $M$ should be associated with a light neutrino mass $m_{\nu }\sim
10^{-3}$ eV.

Here, we have followed the working hypothesis of \cite{FriemanHillStebWaga}
which states that the effective vacuum energy will be dominated by the
heaviest fields still evolving towards their true minimum. We assume that
the super light PGNB field $\Phi$, with associated mass of order $\sim
m_{\nu }^{2}/f$, will be the last field still rolling down its potential.
Thus we have chosen by hand the constant in eq.(\ref{PGNBeffpot}) so that
when $\Phi $ reaches the minimum it will have zero cosmological constant
associated with it. This choice is essentially a choice of the zero of
energy at asymptotically late times.

The finite temperature behavior associated with these models is extremely
interesting\cite{softboson}. For $N\geq 3$ the $\Phi $ dependent part of the
potential can be written as 
\begin{equation}
c\left( T\right) M^{4}\cos \frac{N\Phi }{f},  \label{finiteT}
\end{equation}
where $c\left( T\right) $ vanishes at high temperature $T$. Thus the high
temperature phase of the theory has a non-linearly realized $U\left(
1\right) $ symmetry where the $\Phi $ potential becomes exactly flat with
value $M^{4}$. Since $M\sim m_{\nu }$ this cosmological constant
contribution will have no effect during nucleosynthesis and through the
matter dominated phase until $T\sim M$. At this time $c\left( T\right) $
reaches its asymptotic value of unity and we have the potential in eq.(\ref
{PGNBeffpot}). For $N=2$, $c\left( T\right) $ changes sign continously,
passing through zero at the critical temperature such that the high
temperature minima become the low temperature maxima and vice-versa. There
is a $Z_{2}$ symmetry in both the low and high temperature phases.

The potential in eq.(\ref{PGNBeffpot}) has regions of spinodal instability,
i.e. where the effective mass squared is negative. These occur when $\cos
N\Phi /f>0.$ If $\Phi $ is in this region, modes of sufficiently small
comoving wavenumber follow an equation of motion that at least for early
times is that of an inverted harmonic oscillator. This instability will then
drive the non-perturbative growth of quantum fluctuations until they reach
the spinodal line where $\cos N\Phi /f=0$\cite{Spinodal}. Since the quantum
fluctuations grow non-pertrubatively large, we have to resum perturbation
theory to regain sensible behavior and this is done by the Hartree
truncation \cite{Spinodal}. The prescription is to first expand $\Phi $
around its (time dependent) expectation value $\left\langle \Phi \left( \vec{%
x},t\right) \right\rangle \equiv \phi \left( t\right) $ as 
\begin{equation}
\Phi \left( \vec{x},t\right) =\phi \left( t\right) +\psi \left( \vec{x}%
,t\right) ,  \label{fluctexpand}
\end{equation}
where the tadpole condition $\left\langle \psi \left( \vec{x},t\right)
\right\rangle =0$ gives the equation of motion for $\phi \left( t\right) $.
The Hartree approximation involves inserting eq.(\ref{fluctexpand}) into eq.(%
\ref{PGNBeffpot}), expanding the cosines and sines, and then making the
following replacements\cite{Spinodal}: 
\begin{equation}
\cos \frac{N\psi }{f}\longmapsto \left( 1-\frac{N^{2}\left( \psi
^{2}-\left\langle \psi ^{2}\right\rangle \right) }{2f^{2}}\right) \exp -%
\frac{N^{2}\left\langle \psi ^{2}\right\rangle }{2f^{2}},\quad \sin \frac{%
N\psi }{f}\longmapsto \frac{N\psi }{f}\exp -\frac{N^{2}\left\langle \psi
^{2}\right\rangle }{2f^{2}}.  \label{hartreereplacements}
\end{equation}
The equations for the field $\phi ,$ and the fluctuation modes $f_{k}$
coupled to the scale factor $a\left( t\right) $ are 
\begin{gather}
\ddot{\phi}+3\frac{\dot{a}}{a}\dot{\phi}-\frac{NM^{4}}{f}\exp -\frac{%
N^{2}\left\langle \psi ^{2}\right\rangle }{2f^{2}}\sin \frac{N\phi }{f}=0,
\label{eqofmotion} \\
\ddot{f}_{k}+3\frac{\dot{a}}{a}\dot{f}_{k}+\left( \frac{k^{2}}{a^{2}}-\frac{%
N^{2}M^{4}}{f^{2}}\exp -\frac{N^{2}\left\langle \psi ^{2}\right\rangle }{%
2f^{2}}\cos \frac{N\phi }{f}\right) f_{k}=0,  \nonumber
\end{gather}
with 
\begin{equation}
\left\langle \psi ^{2}\right\rangle =\int \frac{d^{3}k}{\left( 2\pi \right)
^{3}}\left| f_{k}\right| ^{2}.  \label{twoptfunct}
\end{equation}
The effective Friedmann equation for the scale factor is obtained by use of
semiclassical gravity, i.e. by using $\left\langle T_{\mu \nu }\right\rangle 
$ to source the Einstein equations: 
\begin{equation}
\frac{\dot{a}^{2}}{a^{2}}=\frac{8\pi }{3M_{p}^{2}}\left[ \rho _{m}(t)+\frac{1%
}{2}\dot{\phi}^{2}+\frac{1}{2}\langle \dot{\psi}^{2}\rangle +\frac{1}{2a^{2}}%
\left\langle (\vec{\nabla}\psi )^{2}\right\rangle +M^{4}\left( 1+\cos (N\phi
/f)\exp -\frac{N^{2}\left\langle \psi ^{2}\right\rangle }{2f^{2}}\right) %
\right] ,
\end{equation}
with 
\begin{equation}
\rho _{m}(t)=\rho _{m}(t_{i})\frac{a^{3}(t_{i})}{a^{3}(t)}
\end{equation}
being the matter density and $t_{i}$ being the time at which the PGNB field
begins its evolution.

The interesting feature of the above equations of motion is the appearance
of terms involving $\exp -\frac{N^{2}\left\langle \psi ^{2}\right\rangle }{%
2f^{2}}$. These multiply terms in the potential and its various derivatives
that contain the non-trivial $\phi $ dependence. What we expect to have
happen is that as the spinodally unstable modes grow, they will force $%
\left\langle \psi ^{2}\right\rangle $ to grow as well. This in turn will
rapidly drive the exponential terms to zero, leaving a term proportional to $%
M^{4}$ in the Friedmann equations, which will act as a cosmological constant
at late times.

If we consider the $N\geq 3$ models, then at temperatures larger than $T_{%
\text{crit}}\sim M$ the potential is just given by $M^{4}$ and is swamped by
both the matter and radiation contributions to the energy density. Since the
potential is flat, we expect that the zero mode is equally likely to attain
any value between $0$ and $2\pi $ and in particular, we expect a probability
of order $\sim 1/2$ for the initial value to lie above the spinodal line. If
there was an inflationary period before this phase transition, we expect
that the zero mode will take on the same value throughout the region that
will become the observable universe today.

As the temperature decreases, the non-trivial parts of the potential turn on
and the zero mode begins its evolution towards the minimum once the Universe
is old enough, i.e. $H\left( T_{\text{roll}}\right) \sim m_{\phi }\sim
M^{2}/f$. At the same time, if the zero mode started above the spinodal
line, the fluctuations begin their spinodal growth. Whether the spinodal
instabilities have any cosmological effect depends crucially on a comparison
of time scales, the first being between the time $t_{\ast }$ it takes the
zero mode to reach the spinodal point at $\phi _{\text{spinodal}}/f=\pi /2N$
under the purely classical evolution (i.e. neglecting the fluctuations), and
the time $t_{\text{spinodal}}$ it takes for the fluctuations to sample the
minima of the tree-level potential, so that $N^{2}\langle \psi ^{2}\rangle
/f^{2}\sim {\cal O}\left( 1\right) $. Since the growth of instabilities will
stop at times later than $t_{\ast }$, if spinodal instabilities are to be at
all relevant to the evolution of $\phi $, we need $t_{\text{spinodal}}\ll
t_{\ast }$. By looking at the equations of motion we can argue that\cite
{LargeN} 
\begin{equation}
t_{\text{spinodal}}\simeq \frac{f}{2M^{2}}\ln \frac{f^{2}}{N^{2}\langle \psi
^{2}\rangle \left( t_{i}\right) }+\frac{3}{2H_{i}},  \label{spinodaltime}
\end{equation}
where $t_{i}$ is the time at which the zero mode starts to roll and $H_{i}$
is the Hubble parameter at this time. Furthermore the early time behavior of
the equations of motion gives us 
\begin{equation}
t_{\ast }\simeq \frac{f}{M^{2}}\ln \frac{f}{N\phi (t_{i})}+\frac{3}{2H_{i}}.
\label{classicaltime}
\end{equation}
Comparing eqs.(\ref{spinodaltime},\ref{classicaltime}), we see that to have
the spinodal instabilities be significant we need $\phi ^{2}(t_{i})\ll
\langle \psi ^{2}\rangle \left( t_{i}\right) $.

The other condition that needs to be met is that there should be sufficient
time for the spinodal instabilities to dominate the zero mode evolution
before today. This will ensure that the expansion of the Universe will be
driven by the remnant cosmological constant $M^{4}$ at the times relevant to
the SNIa observations. For large enough initial fluctuations we can make the
spinodal time as early as we need.

What sets the scale of the initial fluctuations? If we assume a previous
inflationary phase, we can treat the PGNB as a minimally coupled massless
field and the standard inflationary results should apply\cite{LindeInf}. The
initial conditions for the mode functions are then given by: 
\begin{equation}
f_{k}\left( t_{i}\right) =\frac{i}{\sqrt{2k^{3}}}H_{DS}\ \text{\ for }\kappa
\leq k\leq H_{i},  \label{longwaveinfic}
\end{equation}
where $\kappa $ is an infrared cutoff corresponding to horizon size during
the De Sitter phase and $H_{DS}$ is the Hubble parameter during inflation.
The short wavelength modes ($k>H_{i}$) have their conformal vacuum initial
conditions. With these initial conditions 
\begin{equation}
\frac{\langle \psi ^{2}\rangle \left( t_{i}\right)}{f^{2}} \simeq \frac{%
H_{DS}^{2}}{4\pi^{2}f^2}\left( N_{\text{e-folds}}-60\right) .
\label{initialinflfluc}
\end{equation}
and for $H_{DS} \simeq 10^{13}$ GeV, $N_{\text{e-folds}}-60\simeq 10^{5}$,
and $f \simeq 10^{15}$ GeV, we would only need that $\phi (t_{i})/f \ll 0.5$%
. These are not outlandish parameter values and we see that very little fine
tuning is required. In fact, there is a great deal of freedom in choosing
the values of these parameters, the only requirement being that the ratio of
parameters appearing in (\ref{initialinflfluc}) not be so small that the
required value of $\phi (t_{i})/f$ is overly restricted.

\begin{figure}[tbp]
\epsfig{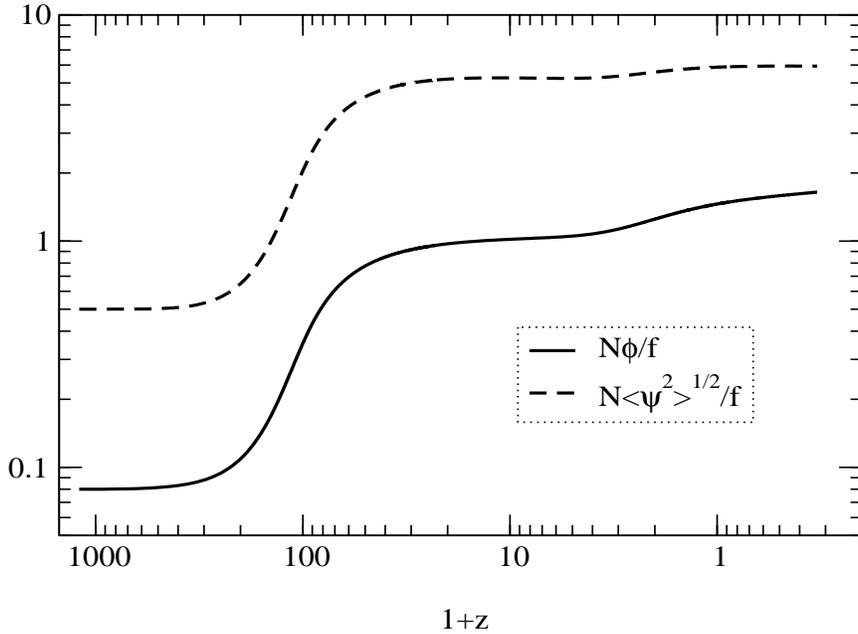}
\caption{The evolution of the zero mode $\protect\phi $ and the root mean
square flucutation $\langle \protect\psi ^{2}\rangle ^{1/2}$ as a function
of the redshift $1+z$ from recombination at $1+z\simeq 1200$ to today at $%
1+z=1$ and beyond. The parameters used for the simulation correspond to 
$M=5.5 \times 10^{-3}$ eV, $f/N=10^{15}$ GeV and 
$(N_{\text{e-folds}}-60)H_{\text{DS}}^{2}=10^{31}(%
\text{GeV})^{2}$. For convenience we have used $(N_{\text{e-folds}}-60)=10$,
which is necessary to avoid numerical implementation of an infrared cutoff
of order $\exp (-10^{5})$. This has the effect of slightly underestimating $%
\langle \protect\psi ^{2}\rangle $ as the evolution proceeds compared to the
case of $(N_{\text{e-folds}}-60)=10^{5}$, but will not change any
qualitative features of the evolution. The initial conditions are $N\protect%
\phi (t_{i})/f=0.08$ and $\dot{\protect\phi}(t_{i})=0$.}
\label{fig1}
\end{figure}

\begin{figure}[tbp]
\epsfig{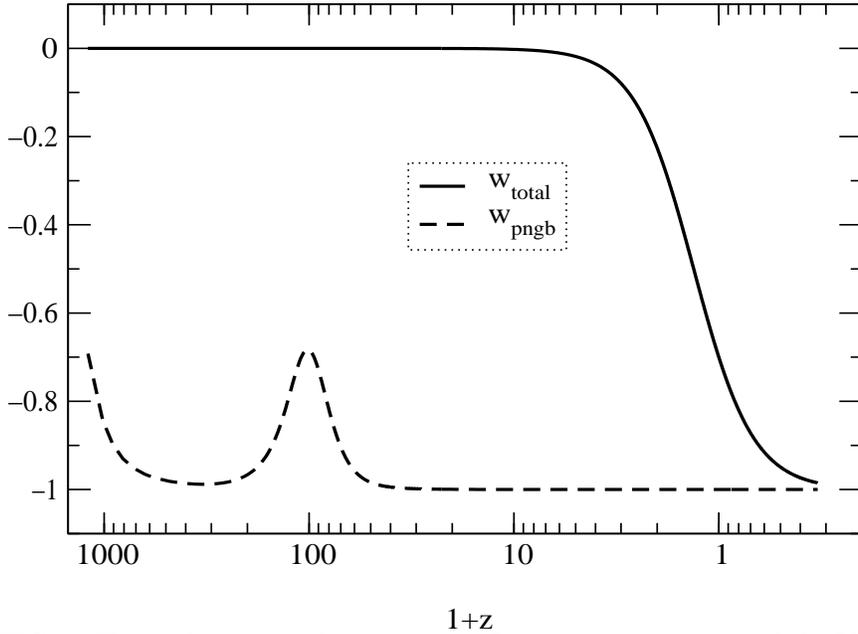}
\caption{The total equation of state $w_{\text{total}} = P_{\text{pngb}}/(%
\protect\rho_{\text{matter}}+\protect\rho_{\text{pngb}})$ and the PNGB
equation of state $w_{\text{pngb}} = P_{\text{pngb}}/\protect\rho_{\text{pngb%
}}$ as a function of redshift. The parameters were chosen as in Fig.\ \ref
{fig1}.}
\label{fig2}
\end{figure}

In the figures below we use these parameters as well as $M\simeq 5.5%
\times 10^{-3}$ eV
corresponding to neutrino masses, beginning the evolution at a redshift $%
1+z=1200$. In Fig.~\ref{fig1} we plot the numerical evolution of the zero
mode and of the growth of the fluctuations, while Fig.~\ref{fig2} shows the
equation of state of the PNGB field and the total equation of state
including PNGB and matter components as a function of redshift. What we
quickly infer from these graphs is that the evolution of the Universe
becomes dominated by the remnant cosmological term, leading to an evolution
toward a late time equation of state $w\equiv P/\rho \simeq -1$. The
equation of state today is seen to be $w \simeq -0.7$ and indicates a matter
component $\Omega_{\text{matter}}=0.3$ and a cosmological constant-like
component $\Omega_{\text{pngb}}=0.7$. Because the PNGB component has an
equation of state $w=-1$ by a redshift as high as $z=50$, these results
reproduce the best fit spatially flat cosmology of the SNIa data\cite{SNIa}.

One feature of this model is that the parameter $M$ is directly related to
the measured value of today's Hubble constant. We find 
\begin{equation}
M=\left( 5.5 \times 10^{-3}\text{eV}\right) \left( 
\frac{H_{0}}{65\frac{\text{km}}{\text{s$\cdot $Mpc}}}\right)^{1/2} 
\left( \frac{\Omega _{\text{pngb}}}{0.7}\right)^{1/4}\;,
\end{equation}
which is to be compared to the observed $90\%$ confidence range of $\Delta
m^{2}$ from the Super Kamiokande contained events analysis\cite{superK1} of $%
5\times 10^{-4}\text{eV}^{2}<\Delta m^{2}<6\times 10^{-3}\text{eV}^{2}$, and
the more recent results of the up-down asymmetry analysis\cite{superK2}
which indicates a range $1.5\times 10^{-3}\text{eV}^{2}<\Delta
m^{2}<1.5\times 10^{-2}\text{eV}^{2}$, whereas the small and large mixing
angle MSW solutions of the solar neutrino problem\cite{MSW,solarnus} yield 
a range of $5\times 10^{-6}\text{eV}^{2}<\Delta m^{2}<4\times 10^{-4}%
\text{eV}^{2}$. 

There is no shortage of models to explain the accelerating expansion of the
universe. However, most options are lacking in motivation and require
significant fine tuning of initial conditions or the introduction of a fine
tuned small scale into the fundamental Lagrangian. We too have a fine tuned
scale: the neutrino mass. However, we can take solace in the fact that this
fine tuning is related to a particle that can be found in the Particle Data
Book\cite{pdb}, with known mechanisms to produce the required value, and
experiments dedicated to its measurement.

The model itself is also relatively benign, not requiring invocations of
String or M-theory to justify its potential. Chiral symmetry breaking
leading to PNGB's is not unheard of in nature (pions do exist after all!),
and should probably be expected in GUT or SUSY symmetry breaking phase
transitions involving coupled scalars. This, together with the dynamical
effects of backreaction allow the present model to be successful in
explaining the data with only minor tuning of initial conditions.

\bigskip

\begin{acknowledgement}
\bigskip D.C. was supported by a Humboldt Fellowship while R.H. was
supported in part by the Department of Energy Contract DE-FG02-91-ER40682.
\end{acknowledgement}

\newpage

\end{document}